%% file: main.tex
\documentclass[sigconf]{acmart}

\AtBeginDocument{%
  \providecommand\BibTeX{{%
    \normalfont B\kern-0.5em{\scshape i\kern-0.25em b}\kern-0.8em\TeX}}}

\setcopyright{acmlicensed}
\copyrightyear{2024}
\acmYear{2024}
\acmDOI{10.1145/3613905.3650937}

\acmConference[CHI '24]{CHI EA '24: In Extended Abstracts of the CHI Conference on Human Factors in Computing Systems}{May 11--16, 2024}{Honolulu, HI}
\acmBooktitle{CHI EA '24: In Extended Abstracts of the CHI Conference on Human Factors in Computing Systems,
 May 11--16, 2024, Honolulu, HI} 
\acmISBN{979-8-4007-0331-7/24/05}

\usepackage[frozencache,cachedir=.]{minted}
\newmintinline[inlinecode]{python}{
    bgcolor=green!10,
    fontsize=\normalsize
}

\begin{document}

\title{Exploring How Multiple Levels of GPT-Generated Programming Hints Support or Disappoint Novices}

\author{Ruiwei Xiao}
\orcid{0000-0002-6461-7611}
\email{ruiweix@andrew.cmu.edu}
\affiliation{%
  \institution{Carnegie Mellon University}
  \city{Pittsburgh}
  \state{Pennsylvania}
  \country{USA}
}

\author{Xinying Hou}
\orcid{0000-0002-1182-5839}
\email{xyhou@umich.edu}
\affiliation{%
  \institution{University of Michigan}
  \city{Ann Arbor}
  \state{Michigan}
  \country{USA}
}

\author{John Stamper}
\orcid{0000-0002-2291-1468}
\email{jstamper@cs.cmu.edu}
\affiliation{%
  \institution{Carnegie Mellon University}
  \city{Pittsburgh}
  \state{Pennsylvania}
  \country{USA}
}

\begin{abstract}
  \input{sections/00_abstract}
\end{abstract}

\ccsdesc[500]{Human-centered computing~Empirical studies in HCI}

\keywords{Programming Hint, Large Language Model, GPT, Introductory Programming, Help-seeking}



\maketitle

\section{Introduction}
  \input{sections/01_Introduction}

\section{Related Work}
  \input{sections/02_related_work}

\section{LLM Hint Factory}
  \input{sections/03_Method}

\section{Method}
  \input{sections/04_Experiment}

\section{Results}
  \input{sections/05_Results}

\section{LIMITATIONS}
  \input{sections/06_Limitation}

\section{Discussion}
  \input{sections/07_Discussion}

\section{Conclusion}
  \input{sections/08_Conclusion}

\bibliographystyle{ACM-Reference-Format}
\bibliography{citations}

\clearpage
\onecolumn
\appendix
    \section{Prompt for each level of hint}
        \input{sections/09_Appendix}
        
    \section{Example of good and bad hints for each criteria}
        \input{sections/10_good_n_bad_hints}

\clearpage
\end{document}

%% file: sections/00_abstract.tex

Recent studies have integrated large language models (LLMs) into diverse educational contexts, including providing adaptive programming hints, a type of feedback focuses on helping students move forward during problem-solving. However, most existing LLM-based hint systems are limited to one single hint type. To investigate whether and how different levels of hints can support students' problem-solving and learning, we conducted a think-aloud study with 12 novices using the LLM Hint Factory, a system providing four levels of hints from general natural language guidance to concrete code assistance, varying in format and granularity. We discovered that high-level natural language hints alone can be helpless or even misleading, especially when addressing next-step or syntax-related help requests. Adding lower-level hints, like code examples with in-line comments, can better support students. The findings open up future work on customizing help responses from content, format, and granularity levels to accurately identify and meet students' learning needs. 

%% file: sections/01_Introduction.tex
Programming novices need in-time, effective support when getting stuck. Intelligent programming tutors (IPTs) can generate adaptive feedback, which can be a scalable solution to beginners’ high demands for help. In the past two years, LLMs have been demonstrated to be effective in generating code fixes and explanations, leading to their application in programming tutors for generating hints to students \cite{liffiton_codehelp_2023}. Current LLM-based systems mainly focus on leveraging the language model’s advantage to provide hints as formative feedback in more descriptive, high-level natural language, imitating the human teaching assistant's tone \cite{kumar_quickta_nodate}. 

Despite the benefits of providing multiple levels of hints in intelligent tutoring systems studies \cite{aleven_2016_help}, few studies investigated the diversity in LLM-generated programming hints or how to deliver these hints to maximize student learning. To bridge these gaps, we leveraged existing literature on how to deliver formative feedback \cite{shute_focus_2008,stamper_experimental_2013} to develop the LLM Hint Factory, which provides 4 different levels of LLM-generated hints in a thread for each hint request, from general natural language guidance to concrete code snippet suggestion. We then conducted a user study with 12 novice programming learners to complete three programming learning tasks with pre- and post-test using the LLM Hint Factory. Results showed that, delivering high-level natural language hints alone can be insufficient, or sometimes even harmful to students, especially when students are confused about next-step logic or syntax-related details. Adding low-level hints, particularly example code pieces with in-line comments can optimally help students in most cases. This work provides a novel contribution by designing the LLM Hint Factory, a system to provide scalable, high-quality programming hints at various levels, and revealing patterns of appropriate hint levels for varied help-seeking circumstances and the corresponding reasons. This work also contributes to the field by highlighting the need to distinguish students' demands based on their help-seeking contexts. The response to students' help requests should be personalized in terms of content, format, and granularity to meet students' diverse needs. It also opens up work into what and how future instructional agents, including educational chatbots and AI teaching assistants, should be designed to respond to students' diverse help requests optimally.


%% file: sections/02_related_work.tex
\subsection{Hint delivery in traditional intelligent programming tutors}
Hint is a special type of formative feedback. Hints provided in programming problem-solving mostly focus on providing information to guide learners towards the next steps for a correct solution \cite{phdthesis}. In traditional IPT research, different levels of hints have been implemented with theoretical support and examined by empirical evidence. Hints in some systems are designed in high-level natural language with limited programming syntax to simulate human tutors. For instance, hints from a rule-based IPT Lisp tutor are provided in human tutors' tone to explain how to do next ~\cite{anderson_lisp_1985}. Another system, ITAP, is a data-driven Python programming tutor that also provides one-sentence next-step suggestions in natural language with location, value, action, and context information ~\cite{rivers_data-driven_2017}. 

In contrast to providing natural language hints, other tutoring systems provide more comprehensive and concrete support, such as worked example code (a code solution to a similar problem) and bottom-out code (the code solution to this problem) hints to reduce students' cognitive load. For example, iSnap ~\cite{price_isnap_2017} presents the student’s buggy part of the code with suggested next-step code side by side to demonstrate what to do next. In addition, Yahoo! Pipes \cite{kuttal_debugging_2013} provides external links containing worked examples to learners as hints. What's more, programming tutors that provide on-demand Parsons problems related to the current question can be content-wise seen as button-out code hints when students seek for certain syntax on one of the lines or finish solving the Parsons problem \cite{hou_using_2022, hou_2023_understanding}. Moving from single hint type to multi-levels of hints, \citeauthor{suzuki_exploring_2017} identified 5 types of hints through program synthesis \cite{suzuki_exploring_2017}. Providing multiple levels of hints for one request not only provides learners with extended sources of help, but also allow them to compare the helpfulness of different levels of hints under different conditions. For example, Hint Factory \cite{stamper_hint_nodate} derives a hint sequence into 4 parts: three of them focus on one of the certain components in the question structure and one bottom-out hint on the explicit action needed to proceed to the next step. What's more, Ask-Elle \cite{gerdes_ask-elle_2017} organized descriptive and worked-out code hints of the next step for all potential problem-solving strategies. 

Regardless of the varied hint designs and their promising results to support student learning, scalability remains a key challenge across all levels of hint generation in traditional IPT systems. Conventionally, delivering adaptive hints either require taxing expert inputs (e.g. annotating answers \cite{gerdes_ask-elle_2017}, defining rules ~\cite{anderson_lisp_1985}, etc.), or need to be driven by a high volume, comprehensive dataset \cite{rivers_data-driven_2017, suzuki_exploring_2017}. The emergence of LLMs provides a new opportunity to automate the hint generation process and customize the hints to target students' diverse needs.

\subsection{LLM-Based Intelligent Programming Tutors}
After recognizing the ability of LLMs to generate code fixes \cite{macneil_experiences_2023, sahai_improving_nodate}, explanations \cite{chen2023gptutor}, and pedagogical conversations \cite{tack_ai_2022}, significant effort has been invested in the creation and evaluation of LLM-based intelligent programming tutors with diverse content focus and granularity. Most existing studies are dedicated to providing higher-level hints in paragraphs of natural language and cautiously avoid providing concrete solution-like code to prevent overuse or over-reliance. However, some of these studies acknowledged that students prefer to have hints with varied levels of details than high-level ones alone \cite{roest_automated_nodate,liffiton_codehelp_2023}. Meanwhile, \citeauthor{kazemitabaar_studying_2023}  \cite{kazemitabaar_studying_2023} pointed out that having access to the AI code generator and generated code snippets did not impede learning gains. Therefore, we believe that code examples for similar problems have the potential to be incorporated as a more specific level of hints.
Among the recent deployment of LLM-based IPTs in intro CS education, to our knowledge, no work has explored the effects of different hint levels on student problem-solving. We identified only one study that provided three levels of code explanation (line-by-line, summary, and key concepts) to students side-by-side in an E-book \cite{macneil_experiences_2023}. However, explanations in this system are additional instructional explanations for instructional materials rather than hint feedback to guide students to move forward in problem solving. Our work distinguishes itself from existing approaches by comparing the effects between different levels of next-step hints on supporting novices' problem-solving. More specifically, our system (the LLM Hint Factory) provides four levels of hints for each hint request, giving students full autonomy to use any hint level and ask for hints whenever needed. Hint level design follows principles and definitions in previous formative feedback work with an emphasis on programming exercises\cite{keuning_systematic_2019}. 


%% file: sections/03_Method.tex
\begin{figure*}[ht]
    \includegraphics[width=1.00\textwidth]{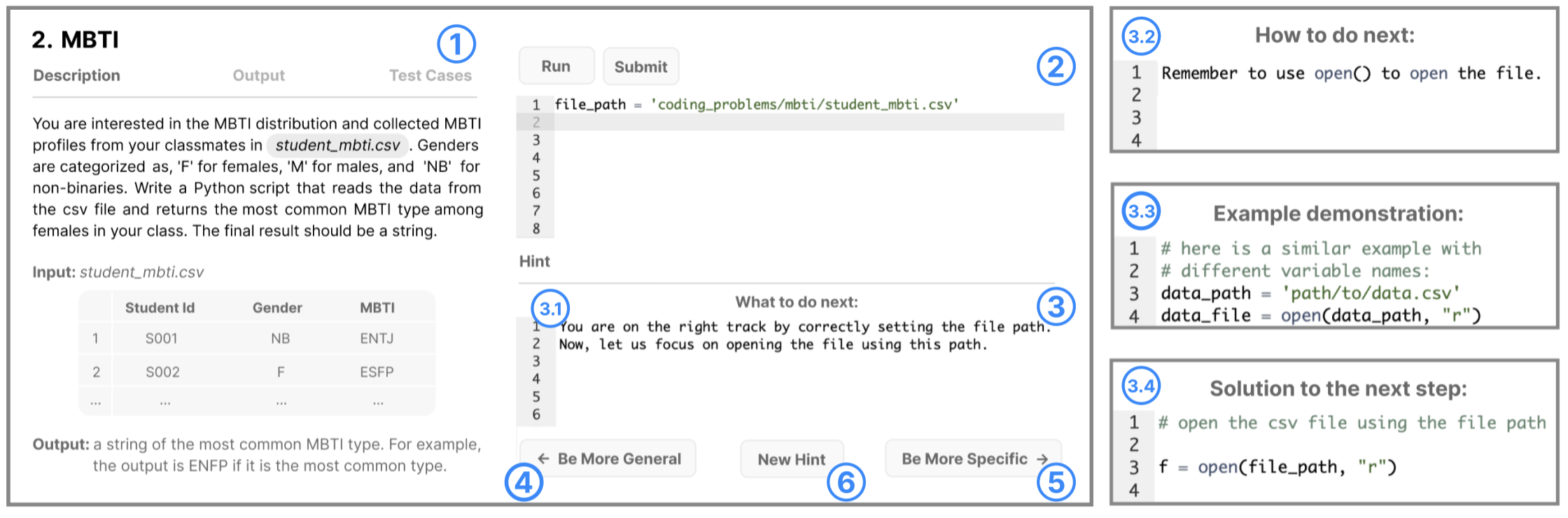}
    \centering
    \caption{LLM Hint Factory Interface. (1) Problem description; (2) Code editor; (3) Hint Section; (3.1) Orientation Hint: the 1st level hint, informs students where they should focus; (3.2) Instrumental Hint: the 2nd level hint, informs students how to do next in concise, descriptive sentences; (3.3) Worked Example Hint: the 3rd level hint, shows students an example code snippet that is similar to the code they need to write for their next step to solve the current problem; (3.4) Bottom-Out Hint: the 4th level hint, shows students the exact code they need to write for the next step to solve the current problem; (4) Click "Be More General" button to see the previous level's hint; (5) Click "Be More Specific" to see the next level's hint; (6) Click "New Hint" to generate a new set of hint.}
  \label{fig:interface}
\end{figure*}

Based on our goals, we designed and developed the LLM Hint Factory, a system that provides four levels of hints, from the most high-level natural language guidance to the most specific code snippets, to CS1 students in real-time when they ask for help in solving a programming task (Figure \ref{fig:interface}). 

\textbf{Hint Design in the LLM Hint Factory} Programming hints in the LLM Hint Factory are formative, next-step feedback within an economy of words \cite{denny_designing_2021}. Compared to many existing LLM-based IPTs, which provide long explanations mixed with small snippets of code trying to resolve all issues in the student's code, our feedback is segmented into individual steps to break the information into manageable chunks. By doing this, we aim to help learners manage their essential processing in working memory to achieve better learning. Given that even the bottom-out hints only contain part of the solution, this can also prevent the potential over-reliance by showing students a full LLM-generated answer \cite{kazemitabaar_how_2023}. Specifically, we designed four levels of hints, from the highest level to the most specific level: orientation hint, instrumental hint, worked example hint, and bottom-out hint. The division of levels is originally inspired by the hint-level design and definition in Cognitive Tutor \cite{koedinger_exploring_2007}, and we further grounded the definition into programming context and added one more level, worked example hint into the LLM Hint Factory. Prompts for different levels' hints generation can be found in Appendix A.

\textbf{LLM Hint Factory Interface} LLM Hint Factory shares great similarities to a typical online programming environment (Figure \ref{fig:interface}). The left panel includes the problem description and a clickable link to check the input data; the right panel contains a code editor and a "Hint” tab. Students can click “New Hint” to generate a set of four hints based on the current question information and code state, and click “Be More Specific” to read a more explicit hint targeting the same sub-step or click "Be More General" to navigate to the last hint level. The button-triggering for more specific level of hint design adheres to scaffolding guidelines in learning sciences research \cite{malik2017revisiting}, which suggest to provide just right level of assistance needed by each individual.

%% file: sections/04_Experiment.tex
To understand the effectiveness of diverse levels of programming hints, we conducted an IRB (Institutional Review Board)-approved think-aloud study with a pre-test and post-test to investigate how these hints could help students with programming problem-solving and contribute to learning.

\textbf{Participants}
Twelve undergraduate and graduate students were recruited from both a private research institution and a public university in the United States. These students are currently enrolled in or have only taken introductory computer science classes for non-CS majors. To verify the eligibility of the participants, we implemented a screening question designed to confirm that each participant has only taken the CS1 class using Python. 

\textbf{Study Procedure}
The think-aloud session started after checking the students’ ages. During this one-hour session, participants first received a 5-minute introduction about the study consent, experiment procedure, and interface. Then participants spent 10 minutes on the pre-test, 30 minutes on the learning session, and 10 minutes on the post-test. Finally, participants received a 5-minute structured interview about their experiences using the multiple levels of hints (3.1 to 3.4 in Figure \ref{fig:interface}) in the LLM Hint Factory. The interviews aimed at understanding students’ opinions on hints they received, such as whether they were comprehensible and helpful, and their suggestions for future improvement.

\textbf{Experiment Materials}
Three programming tasks (Figure \ref{fig:interface} demonstrates one of these tasks) about data analysis in Python were designed as practice problems for the learning session. To guarantee that the questions and their difficulty levels were suitable for our participants, we adapted them from the homework materials from the two classes from which we recruited, thus participants would be more likely to have the necessary prior knowledge and manage to complete tasks in the given time. The pre-test included eight test questions (1 open-ended, 3 multiple-choice questions, and 4 code-writing questions) about basic data extraction. The post-test contains identical programming questions as the pre-test. 


%% file: sections/05_Results.tex
\subsection{Overall Hint Quality in the LLM Hint Factory}
\subsubsection{LLM Hint Factory can generate high-quality multiple levels of hints}
We first conducted an expert evaluation of the hint quality generated during the user study. We defined a coding rubric to evaluate the quality of generated hints from 6 dimensions (table \ref{tab:rubric}). The rubric was designed based on previous formative feedback design studies \cite{shute_focus_2008,crow_intelligent_2018} and existing works on LLM-generated programming hint evaluation \cite{roest_automated_nodate, nguyen_towards_2022}. Following this coding rubric, two expert raters randomly sampled 20\% of hint requests (resulting in 304 unique hints) generated during the user study to calculate the inter-rater reliability. Due to the high prevalence of the “Yes” label and binary classes, we applied the percent agreement approach and achieved a high agreement across dimensions.
Next, two raters met and resolved the disagreements in their codes and refined the rubric. Then, one rater coded the rest of the data. Besides the manual coding on the first five criteria, the average length of each type of hint was aggregated automatically. The full evaluation results are in Table \ref{tab:expert_result}, and example hints fall into each category can be found in Appendix B.

\begin{table*}[ht]
\fontsize{8pt}{10pt}\selectfont
\caption{The expert evaluation rubric across six categories: Appropriate, Targeted, Comprehensible, Encouragement, Alignment and Length. While Alignment evaluate the interrelationship among a set of four hints, other criteria target per hint level.}
\label{tab:rubric}
\begin{tabular}{ll}
\hline \textbf{Category} & \multicolumn{1}{c}{\textbf{Definition}}                                                                                                     \\ \hline
\textbf{Appropriate - \textit{(Yes/No)}} & Is the hint a suitable next step to construct a solution given the learners’ current code and its level?\\
\textbf{Target - \textit{(Yes/No)}} & Does the generated hint target the required level? \\
\textbf{Comprehensible - \textit{(Yes/No)}} & Is the generated hint understandable? \\
\textbf{Encouragement - \textit{(Yes/No)}} & Does the generated hint provide any form of encouragement? \\ 
\textbf{Alignment - \textit{(0-1)}} & Does the content of each level of hint align with other levels?\\&
0: none of them align with each other; 0.25: only two of them align with each other.\\&
0.5: pairs align with each other; 0.75: three of them align with each other; 1: all hints are well-aligned.\\
\textbf{Length} & For orientation and instrumental hints, what are the number of sentences\\ &  and number of words in each hint? For worked examples and bottom-out \\ & hints, what are the lines of code in each hint? \\

\hline
\end{tabular}
\end{table*}

\begin{table*}[ht]
\fontsize{8pt}{10pt}\selectfont
  \caption{The expert evaluation results on Appropriate, Targeted, Comprehensible and Length for each level of hints.}
  \label{tab:expert_result}
\centering
\begin{tabular}{lcccc}
\hline
\textbf{Type}         & \textbf{Appropriate} & \textbf{Targeted} & \textbf{Comprehensible} & \textbf{Length} \\ \hline
\textbf{Overall}      & 81.91\%             & 99.67\%          & 98.36\%                & -               \\
\textbf{Orientational} & 88.16\%             & 100\%            & 98.68\%                & 30 words; 2 sentences \\
\textbf{Instructional} & 84.21\%             & 98.68\%          & 100\%                  & 29 words; 2 sentences \\
\textbf{Worked-Example} & 75\%               & 100\%            & 100\%                  & 12 lines of code       \\
\textbf{Bottom-Out}    & 78.95\%             & 98.68\%          & 93.42\%                & 7 lines of code        \\ \hline
\end{tabular}
\end{table*}

Overall, all four levels of hints generated by the LLM Hint Factory are \textit{comprehensible} and \textit{well-targeted} to the prompt instruction at the correct level of detail. For the appropriate category, 88.16\% of the orientation hints and 84.21\% of the instrumental hints can point out and elaborate one problem at a time, and for the lower level hints, 75\% of the worked example hints and 78.95\% of the bottom-out hints can elaborate on the previous two levels of natural language hints and response with Python code pieces. However, we observed that sometimes the lower-level hints (worked example and bottom-out hints) contain code for more than one step, such as correcting the import library part and the for loop in one single hint, which was considered inappropriate in this category.
Regarding the \textit{alignment} of the four levels of hints, the average alignment among 4 levels of hints in each hint request is 78.62\%, which means that at least three hints in one request focus on the same problem. According to \textit{length}, for the first two levels of natural language hints, most of them are around 30 words in two sentences. For the remaining two levels of hints (code hints), their average lengths are 12 and 7 lines respectively, with in-line code explanation comments (Table \ref{tab:expert_result}).

\subsubsection{High-quality hints are not always helpful for students’ problem-solving}
We next looked at student interaction logs and recordings to further unpack the effectiveness of hints in guiding students to take correct actions. We found that, out of the expert-labeled high-quality hints, 77.78\% helped students move forward on the right track in problem-solving.

\textit{So what happened to those unresolved help requests?} We conducted thematic coding on students' log data and interview transcripts, and the results showed that 1.43\% hints were misunderstood by the students and therefore led to negative effects. For example, after reading one worked example hint, P2 copied the code \inlinecode{gender = row[0]} without noticing the comment saying that, \inlinecode{# Assuming each row has 'gender' in the first column }, which introduced a wrong index error to their code. The remaining 27.14\% sets of hints did not have an observable impact on students’ progression. For these correct hints, some of them lack visual aids to draw students' attention (e.g. the line of code to \inlinecode{import csv} in the worked example has been missed by three participants). Some other ones were answering questions that misaligned with learners’ questions in mind, making the comprehension process harder. 

Students’ low motivation to learn and high anxiety levels might also impact the forage of critical information in hints. These learners' urge for a direct solution might hinder their comprehension in understanding their current problem. This example demonstrated how one hint is helpful for P10 but not as helpful for P2. When they received the same orientation hint \textit{[Now, let's focus on filtering the data for females before counting MBTI types]}, P10 skimmed the problem description again and said \textit{"Oh, I misread the question, I should only count the most common MBTI type for females"}, and added the \inlinecode{if} statement in the code for the desired target. In contrast, P2 could not identify the error even after reading the hint aloud and never referred back to the problem description. Based on P2’s frequent request for hints even before reading the question prompt, the low learning motivation and high anxiety on task completion may hinder the information forage from hints.

\subsection{The Effectiveness of Different Levels of Programming Hints for Supporting Novices}
\subsubsection{Providing hints till the level of worked example can assist students properly.} 
To understand the impact of different levels of hints, we looked at both student progress during learning sessions and their changes from pre-test to post-test. Recall that students received four levels of hints for each hint request, and the expected behavior of using a hint is to guide students toward correct changes. Therefore, we started with investigating the hint level that students needed to reach to modify the code correctly during the learning session. Among those hints labeled as appropriate by experts, in 35 (59.32\%) of the instances, students were able to make correct code changes after students saw the worked-example hints; and in 5 instances, students correctly modified code after receiving bottom-out hints. In addition, while students generally revealed positive reactions (e.g. saying \textit{"that makes sense"}) to the high-level hints, lower-level example code hints were more likely to lead to correct programming actions. For example, P24 requested a hint because she forgot the syntax of a for loop. When seeing the instrumental hint, she said \textit{"Yeah which is exactly where I’m stuck at, but I want to know the exact syntax"} and immediately moved on to the example code hint. When reading through the example code in this hint, she stated, \textit{"Oh yeah, true, looks like the correct syntax should not be the curve bracket. I should use the column."} Then they changed to the correct syntax, said \textit{"...I wanna stop here and try to run again and see..."}, and did not proceed to the bottom-out hint.

As pre- and post-test shared the same questions, we also compared student answers from pre- to post-test to investigate how different levels of hints played a positive role in assisting students' expansion and refinement of their programming knowledge. For example, P5 could not write \inlinecode{import csv} in the pretest, struggled with the same item during learning session until seeing a worked example hint, and wrote this line correctly in the post-test. We found that, for those who made positive changes from pre- to post-test, all of them corrected their misconception or added new knowledge after absorbing the content in the last two levels of hints. Specifically, 75\% of them were satisfied with the worked example hints and did need to get to the bottom-out level, while 25\% still needed to get to the bottom-out level to make meaningful changes. Deriving from both hint request data and pre-post tests data, we deduced that hints up to the level of a worked example code can deliver proper help, primarily due to its frequent effectiveness and its balance between specificity, surpassing the high-level hints. Also, compared to bottom-out code hints, applying a worked example code hint requires higher cognitive engagement, which can encourage more meaningful learning.

\subsubsection{Worked example hints provide comprehensive help on next-step and syntax-related hint requests}
\begin{figure}[ht]
  \centering
  \includegraphics[width=\linewidth]{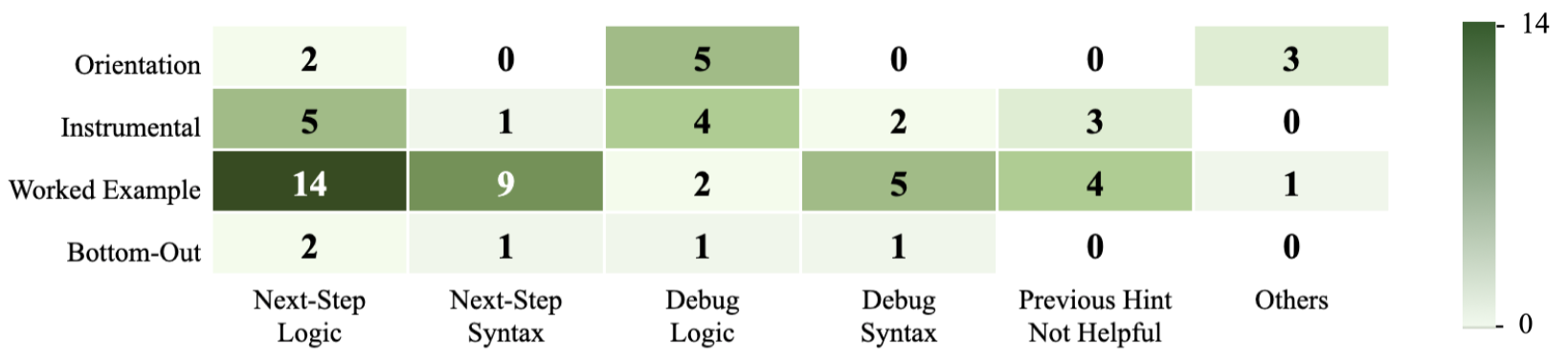}
  \caption{Number of effective level of hint for each help-seeking type. Most of requests related to next-step (NL, NS) and syntax (NS, DS) are will be resolved when learners utilized worked example hints. Most of DL requests can be answered by high-level hints.}
  \label{fig:heatmap}
\end{figure}

To further examine the potential circumstances when each level of hint could be helpful, we contextualized learners’ reasons to request help in programming learning as six types as shown in Table \ref{tab:help_seeking_types}, following the help-seeking model \cite{roll_improving_2011}. After categorizing motivations of hint requests, we mapped effective levels of hints by the type of requests (Figure \ref{fig:heatmap}). Our results showed that, when students’ confusions pertained to the progression to the next step or were syntax-related (the request types NL, NS, DS), most of their correct actions have not been taken until they saw the worked example hints. Drawing from these observations, it is recommended to furnish students with exemplar code blocks, particularly when their confusions are related to the progress to the next-step (NL, NS requests) or programming syntax (DL, DS requests).  

\begin{table*}[ht]
\fontsize{8pt}{10pt}\selectfont
\caption{Six help-seeking types during programming problem-solving process. Adapted from the help-seeking model \cite{roll_improving_2011}.}
\label{tab:help_seeking_types}
\begin{tabular}{ll}
\hline \textbf{Type} & \multicolumn{1}{c}{\textbf{Definition}}                                                                                                     \\ \hline

Next-Step Logic (NL) & When a student cannot figure out the logic of the next step.\\
Next-Step Syntax (NS) & When a student knows the logic of the next step but do not know the syntax.\\
Debug Logic (DL) & When student wants to debug code with errors in logic, such as the \\& misunderstanding of problem description.\\
Debug Syntax (DS) & When student wants to debug code with errors in syntax like wrong indentation.\\
Previous Hint Not Helpful (PNH) & When a student request for a new hint because the previous hint is not helpful.\\
Others & When a student asks for hints without a clear purpose, in our context, this type \\& of request happens when learners are curious about this feature and try it out.\\

\hline
\end{tabular}
\end{table*}

When being asked about why the worked example hints are helpful for next-step or syntax-related requests, participants highlighted that applying the help from the worked example is straightforward, code-to-code level of help, while high-level hints require more tacit knowledge such as translating natural language into programming language or code structure visualization. P2 thought the example hint was useful as, \textit{"when I had like, a little bit of confusion like I think I knew what I had a general idea from high-level hints what I was supposed to do but not didn't know how to execute it"}. In another case, when P8 cannot tell when to write a \inlinecode{return} and when not to, she did not realize there is no function definition in her code when reading the instrumental hint, \textit{[return should be inside of a function structure]}. However, she found this part out immediately when she compared her code with the worked example hint and saw the function structure in the worked example was not in her code.

\subsubsection{High-level hints can provide more concise help for students to debug logic errors}
For requests about logic debugging (DL), providing high-level (orientation, instrumental) natural language guidance is often sufficient for students to effectively address and resolve errors. Multiple participants (n=4) have been reminded by the first two levels' hints to include data for females only as the problem description required, and one participant corrected the values they store for each company after seeing \textit{[The first error is that you're not updating the total launches for each company. Both successful and unsuccessful launches should be counted in the total.]}

\subsubsection{High-level hints can cause more misunderstanding and frustration than low-level hints}
33.3\% of the students (n=4) reported their frustration when the high-level hints repeated things they already know. For example, P5 expressed as \textit{"I got a little bit frustrated because... It was telling me something I already knew."}. Moreover, high-level hints can be easily misinterpreted due to their unspecified, descriptive nature. When P10 saw the orientation hint \textit{[This is a really good start, you're on the right track with opening the file. Now, let's focus on reading the CSV file into a 2-D list.]}, they misunderstood the 2-D list as the output data structure of the final result, while it actually means the input data structure. Such misunderstanding was not clarified until they saw the code in the worked example hint. Additionally, high-level erroneous hints are harder to identify than low-level ones since errors in code blocks are easier to compare and more concrete. In this study, all three misleading hints that caused incorrect actions were high-level ones, while in every instance where participants received a buggy low-level hint, they were able to identify the errors.

%% file: sections/06_Limitation.tex
There are three major limitations to this study. The primary limitation to the generalization of these results is the small sample size of 12 programming novices from two research universities and the limited scope of knowledge used in the experiment. More generalizable findings need to be drawn from larger and more diverse participant populations with more different CS1 programming problems. Also, empirical use cases of the system can vary from experiment settings, and further classroom studies need to be conducted to further evaluate the effectiveness of hints. Lastly, the performance and cost of this system highly rely on models developed by OpenAI. Open-source, domain specific models would make this system more affordable and adjustable for researchers and educational practitioners.

%% file: sections/07_Discussion.tex
In this work, to investigate the effectiveness of multi-level hints during programming problem-solving, we conducted a think-aloud study with 12 novices using the LLM Hint Factory. Our main finding is that, high-level hints are often insufficient to assist learners' requests, and providing code example level hint can assist students more properly. We propose the following design suggestions for our ongoing system improvement and future LLM-based learning systems.

\subsection{Design Suggestion 1: Personalize the help response design under different circumstances}
To avoid the rising concerns of over-reliance on LLM in learning contexts, existing GPT-powered programming tutors are deliberately designed to be natural language focused. The goal for this design is to assist students in concept understanding rather than giving the solution straight away. However, our results suggest that, in most cases, providing high-level hints alone are insufficient to help students, especially on next-step-related and syntax-related help requests. Moreover, the concise nature of the high-level hints could confuse students and cause frustration. This could lose their trust, prompting students to switch to ChatGPT for a full answer. Such findings suggest that the responses to students' help requests should be personalized based on their existing problem-solving stage and prior knowledge level. 

\subsection{Design Suggestion 2: Design semi-structured help-seeking interaction for flexibility and convenience}
Rather than more free-style chatbot interaction, the LLM Hint Factory chose a more structured, button-triggered help-seeking interaction to make hints more accessible by reducing learners’ burden of prompting quality questions, which novices are particularly bad at\cite{liffiton_codehelp_2023}. However, seven participants mentioned in the interview that this lack of flexibility sometimes resulted in misalignment between students’ questions in mind and the problem hints tried to address, which reduced the perceived helpfulness of the hints. In the future, we intend to design a more scaffolded hint request interaction such as including a menu to allow students to select help request type or a grounded abstraction matching approach to enhance learners' question-asking technique \cite{liu_what_2023}. From a learners’ perspective, this interaction is flexible and convenient and potentially helps to develop skills such as debugging and planning. From the system’s perspective, the hints content will be less likely to be misaligned with students, and the level shown to students can also be adapted based on the request type. 


%% file: sections/08_Conclusion.tex
In this paper, we used the LLM Hint Factory to evaluate the effectiveness of multiple levels of hints on student problem-solving and learning. Our main finding is that, unlike offering high-level natural language hints alone, delivering LLM-generated hints to the level of worked example hints can provide a sufficient amount of detail in a straightforward way to students. This can facilitate their thinking and enable them to correctly proceed to the next step more effectively. The findings emphasize the need to customize responses to students' help requests from content, format, and granularity levels to accurately identify and meet student needs in future learning systems.



%% file: sections/09_Appendix.tex
\subsection{Example student's current code state}
\subsubsection{Problem Description} You have an ice cream shop with ordering data stored in orders.csv. Your task is to find the best seller(s) and print it out.
\subsubsection{Student's Code} The following is the code when the student triggered the "New Hint" button for help.
\begin{verbatim}
import csv 
with open('csvIceCreamShop.csv', 'r') as file:
   reader = csv.reader(file)
   data = list(reader)
   for row in data:    
        print(row)
\end{verbatim}

\subsection{Orientation Hint}
\subsubsection{Definition} The hint informs students where they should focus.
\subsubsection{Related Prompt} The next-step hint should include at most one incomplete subgoal without any other information. The response should be within 10-50 words. Do not list the errors and subgoals directly. The rest of the hint should be consistent with the suggestion in this hint. 
\subsubsection{Example} This is a really good start! You've successfully read the data into a list. Now, let's focus on the next subgoal: looping over each line in the data.

\subsection{Instrumental Hint}
\subsubsection{Definition} The hint informs students how to do next in concise, descriptive sentences.
\subsubsection{Related Prompt} Describes the first error and the correct way to do it without providing a specific solution.  The response should be within 10-50 words. 
\subsubsection{Example} The first error is that you are printing each row instead of counting the flavors. You need to loop through each row and update the count of each flavor in a dictionary.

\subsection{Worked Example Hint}
\subsubsection{Definition} The hint shows students an example code snippet that is similar to the code they need to write for their next step to solve the current problem.
\subsubsection{Related Prompt} Provides a similar example code in different scenarios and variable names similar to the bottom-out code. The number of lines and code syntax should be exactly the same as the code in the instrumental hint.
\subsubsection{Example} \# Imagine you have a bookstore and you want to find out which genre is the most popular. You would do something like this:

\begin{verbatim}
for row in data:
    genre = row[1] # Assuming the genre is in the 2nd column
    if genre in genre_counts:
        genre_counts[genre] += 1
    else:
        genre_counts[genre] = 1
\end{verbatim}

\subsection{Bottom-Out Hint}
\subsubsection{Definition} The hint shows students the exact code they need to write for the next step to solve the current problem.
\subsubsection{Related Prompt} Follow the suggestion in the worked example hint, if the student's code has no error and is just incomplete, the hint code should be longer than the student's code. provides the correct version of code on the line(s) first error occurred, or the next substep this student needs to take within 5 lines of code. The code should be organized into an html code block or in an ace editor, and the explanation should be written as in-line comments of code. ALL of this hint should be written in code editor. 
\subsubsection{Example} \# Let's focus on counting the flavors. You'll need to loop through each row and update the dictionary with the count of each flavor.
\begin{verbatim}
# Assuming the flavor is in the first column (index 0)
for row in data:
    flavor = row[0]
    if flavor in flavour_counts:
        flavour_counts[flavor] += 1
    else:
        flavour_counts[flavor] = 1
\end{verbatim}

%% file: sections/10_good_n_bad_hints.tex
\subsection{Appropriate}
\subsubsection{Student's Code}
\begin{verbatim}
import csv
file_path = 'coding_problems/icecream/csvIceCreamShop.csv'
best_sellers = []
f = open(file_path, "r")
reader = csv.reader(f)
data = list(reader)
\end{verbatim}
\subsubsection{Example Appropriate Bottom-Out Hint} When hint gives information within one step:
\begin{verbatim}
data = list(reader)
for row in data:
# iterate through all items in data
\end{verbatim}
\subsubsection{Example Not Appropriate Bottom-Out Hint} When hint gives information for more than one step:
\begin{verbatim}
data = list(reader)
for row in data:
    for flavor in row:
        if flavor in best_sellers:
            best_sellers['flavor'] += 1
        else:
            best_sellers['flavor'] = 0 
\end{verbatim}

\subsection{Target}
\subsubsection{Example Hint Targeting on orientational level} You are moving forward! You successfully read the data, and \textbf{now try to iterate through the items in data}.

\subsection{Comprehensible}
\subsubsection{Example Comprehensible Hint} The first error is that you're not separating the counts by gender. You need to \textbf{have separate counts for each gender to find the most common MBTI type among females}.
\subsubsection{Example Not Comprehensible Hint} This is a really good start! You're on the right track with setting up a dictionary to count the MBTI types. Now, focus on \textbf{how you're updating the counts in the dictionary}.

\subsection{Encouragement}
\subsubsection{Example Encouraging Hint} \textbf{This is a really good start, and you're on the right track with using a dictionary to count the MBTI types.} However, consider how you're currently counting MBTI types for all genders together.
\subsubsection{Example Not Encouraging Hint} Consider how you're currently counting MBTI types for all genders together.

\subsection{Alignment}
Given a worked example hint:
\begin{verbatim}
seats = list(reader)
for row in seats:
    print(row)
\end{verbatim}
\subsubsection{Example Aligned bottom-out Hint} When the bottom-out hint provides consistent amount and content of information than the worked example hint.
\begin{verbatim}
data = list(reader)
for row in data:
    # print row to see the structure and content of each row
\end{verbatim}
\subsubsection{Example Not Aligned bottom-out Hint} When the bottom-out hint provides inconsistent amount or content of information than the worked example hint.
\begin{verbatim}
data = list(reader)
for row in data:
    for flavor in row:
        if flavor in best_sellers:
            best_sellers['flavor'] += 1
        else:
            best_sellers['flavor'] = 0 
\end{verbatim}